\documentclass[a4paper]{spie}  %>>> use this instead for A4 paper
\usepackage[]{graphicx}
\usepackage{amsmath}
\usepackage{amssymb}

\title{Modeling of Closure Phase Measurements with
AMBER/VLTI - Towards Characterization of Exoplanetary Atmospheres
}

\author{Viki Joergens and Andreas Quirrenbach
\skiplinehalf Sterrewacht Leiden / Leiden Observatory, PO Box 9513, NL-2300 RA
Leiden, The Netherlands }

\authorinfo{Further author information: (Send correspondence to V.J.)\\
V.J.: E-mail: viki@strw.leidenuniv.nl}
\begin{document}
  \maketitle

%%%%%%%%%%%%%%%%%%%%%%%%%%%%%%%%%%%%%%%%%%%%%%%%%%%%%%%%%%%%%
\begin{abstract}
Differential phase observations with a near-IR interferometer offer a way to
obtain spectra of extrasolar planets. The method makes use of the wavelength
dependence of the interferometer phase of the planet/star system, which depends
both on the interferometer geometry and on the brightness ratio between the
planet and the star. The differential phase is strongly affected by
instrumental and atmospheric dispersion effects. Difficulties in calibrating
these effects might prevent the application of the differential phase method to
systems with a very high contrast, such as extrasolar planets. A promising
alternative is the use of spectrally resolved closure phases, which are immune
to many of the systematic and random errors affecting the single-baseline
phases. We have modeled the response of the AMBER instrument at the VLTI to
realistic models of known extrasolar planetary systems, taking into account
their theoretical spectra as well as the geometry of the VLTI. We present a
strategy to determine the geometry of the planetary system and the spectrum of
the extrasolar planet from closure phase observations in two steps. We show
that there is a close relation between the nulls in the closure phase and the
nulls in the corresponding single-baseline phases: every second null of a
single-baseline phase is also a null in the closure phase. In particular, the
nulls in the closure phase do not depend on the spectrum but only on the
geometry. Therefore the geometry of the system can be determined by measuring
the nulls in the closure phase, and braking the remaining ambiguity due to the
unknown system orientation by means of observations at different hour angles.
Based on the known geometry, the planet spectrum can then be directly
synthesized from the closure phases.

\end{abstract}

%>>>> Include a list of keywords after the abstract


%%%%%%%%%%%%%%%%%%%%%%%%%%%%%%%%%%%%%%%%%%%%%%%%%%%%%%%%%%%%%
\section{INTRODUCTION}

Since the first extrasolar planet candidate was discovered in 1995 orbiting the
solar-like star 51\,Peg (Mayor \& Queloz 1995), more than 110 extrasolar
planets have been detected by indirect radial velocity surveys (see e.g.\ Mayor
et al.\ 2003 for recent discoveries). In order to learn more about the physics
of these exoplanets and directly detect their atmospheres, observations with
milliarcsecond spatial resolution and at an unprecedented dynamical range are
required (e.g. 10$^{-5}$ for 51Peg). So far, the only observational information
on the spectrum of an extrasolar planet has been obtained for the transiting
planet HD\,209458\,B (e.g.\ Charbonnneau et al.\ 2002, Vidal-Madjar et al.\
2004).

It has been proposed to observe extrasolar planets and their spectra with
ground-based near-IR interferometry through the differential phase or closure
phase method (e.g.\ Quirrenbach \& Mariotti 1997, Akeson \& Swain 1999,
Quirrenbach 2000, Lopez \& Petrov 2000, Segransan 2001, Vannier et al.\ 2004).
An overview of the mathematical basis for the observations of interferometric
phases of extrasolar planets has also been provided by Meisner (2004) and in
particular on the web page referenced therein.

Differential phases are measured between different wavelength bands, which, in
a first order approximation, are subject to the same optical pathlength
fluctuations introduced by atmospheric turbulence. The resulting differential
phase errors can therefore be eliminated in the data reduction. However, higher
order effects due to random dispersion in the atmosphere as well as in the
delay lines (caused mainly by variations of the humidity of the air) can be a
significant source of systematic errors.

A more promising approach is the observation of closure phases, which are
measured for a closed chain of three or more telescopes. The closure phase is
the sum of the individual single-baseline phases of the array and is
independent of atmospheric phase errors. This self-calibration technique was
first recognized by Jennison (1958) and is illustrated in the following
equations. The phase $\Phi'_{ij}$ measured on the baseline between telescope
$i$ and $j$ is the true phase for this baseline $\Phi_{ij}$ plus atmospheric
errors $\psi_i$, $\psi_j$ introduced above the two telescopes plus some random
errors $\epsilon_{ij}$ related to this baseline:

\begin{eqnarray}
\Phi'_{12} = \Phi_{12} + \psi_1 - \psi_2 + \epsilon_{12}  \nonumber\\
\Phi'_{23} = \Phi_{23} + \psi_2 - \psi_3 + \epsilon_{23}  \\
\Phi'_{31} = \Phi_{31} + \psi_3 - \psi_1 + \epsilon_{31}~~. \nonumber
\end{eqnarray}

The closure phase is the sum of the individual observed phases and it
equals the sum of the true phases (plus random non-closing errors):

\begin{eqnarray}
\Phi_\mathrm{cl} = \Phi'_{12} + \Phi'_{23} + \Phi'_{31} =
              \Phi_{12} + \Phi_{23} + \Phi_{31} + (\epsilon_{12}
          + \epsilon_{23}+ \epsilon_{31})~~.
\end{eqnarray}

Closure phase observations from the ground will be possible with the AMBER
(Astronomical Multi-BEam combineR) instrument at the Very Large Telescope
Interferometer (VLTI) operated by ESO. AMBER is a first-generation VLTI
instrument (Petrov et al.\ 2003, Malbet et al.\ 2004) operating in the near-IR
J, H, and K bands from 1.1 to 2.4\,$\mu$m. It is currently undergoing
commissioning at Cerro Paranal and is scheduled to be available for regular
observations in 2005. With a multi-beam combiner element and a fringe
dispersing mode it has the capability of combining the light of three
telescopes (UTs or ATs), and performing spectrally resolved differential and
closure phase observations. There are three different spectral resolutions
available: low ($R=\lambda$/$\Delta \lambda$=35), medium ($R=500\dots1000$),
and high ($R=10000\dots15000$).

%-------------
   \begin{figure}[t]
   \begin{center}
   \begin{tabular}{c}
   \includegraphics[height=9cm]{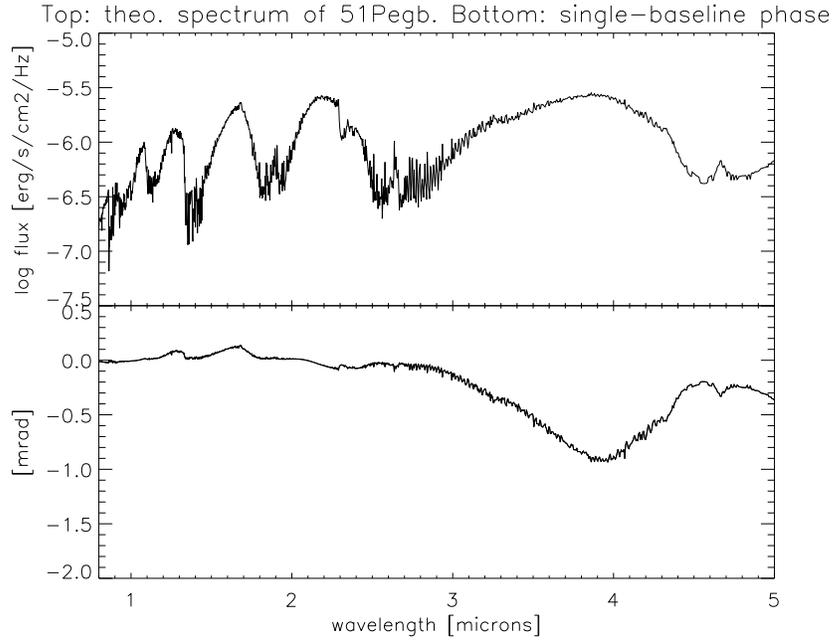}
   \end{tabular}
   \end{center}
   \caption[example]
   { \label{fig:phase}
   {\bf Spectral signal of planet in the single-baseline phase.}
   \emph{Top panel:} Theoretical spectrum of the giant irradiated
planet orbiting the solar-like star 51 Peg (Sudarsky, Burrows \& Hubeny 2003).
Clearly visible are CO and H$_2$O absorption bands in the near-IR. \emph{Bottom
panel:} Single-baseline phases in milliradian from a calculation based on the
theoretical spectrum of 51\,Peg\,B as well as of the host star, and a simulated
observation with the near-IR instrument AMBER/VLTI using the two telescopes UT1
and UT3 (baseline length 102\,m). It can be clearly seen that the spectrally
resolved phases contain spectral information about the planet. We note that due
to atmospheric turbulence, the absolute phase is not an accessible quantity
without a dual-beam facility allowing for phase referencing. For illustration,
we nevertheless display them here. }
   \end{figure}
%-------------

The closure phase signal of a planetary system depends on the interferometer
geometry, the hour angle of the observation, the spectra of the planet and the
star, and on the planetary system geometry. We have modeled the response of the
AMBER instrument at the VLTI for known extrasolar planetary systems, taking
into account their theoretical spectra as well as the geometry of the VLTI. In
the following, we present a method to determine the planetary system geometry
and the planet spectrum from closure phase observations in two steps.

\section{MODELING OF CLOSURE PHASE OBSERVATIONS OF EXOPLANETS
WITH AMBER AT THE VLTI}

%-------------
   \begin{figure}[t]
   \begin{center}
   \begin{tabular}{c}
   \includegraphics[height=9cm]{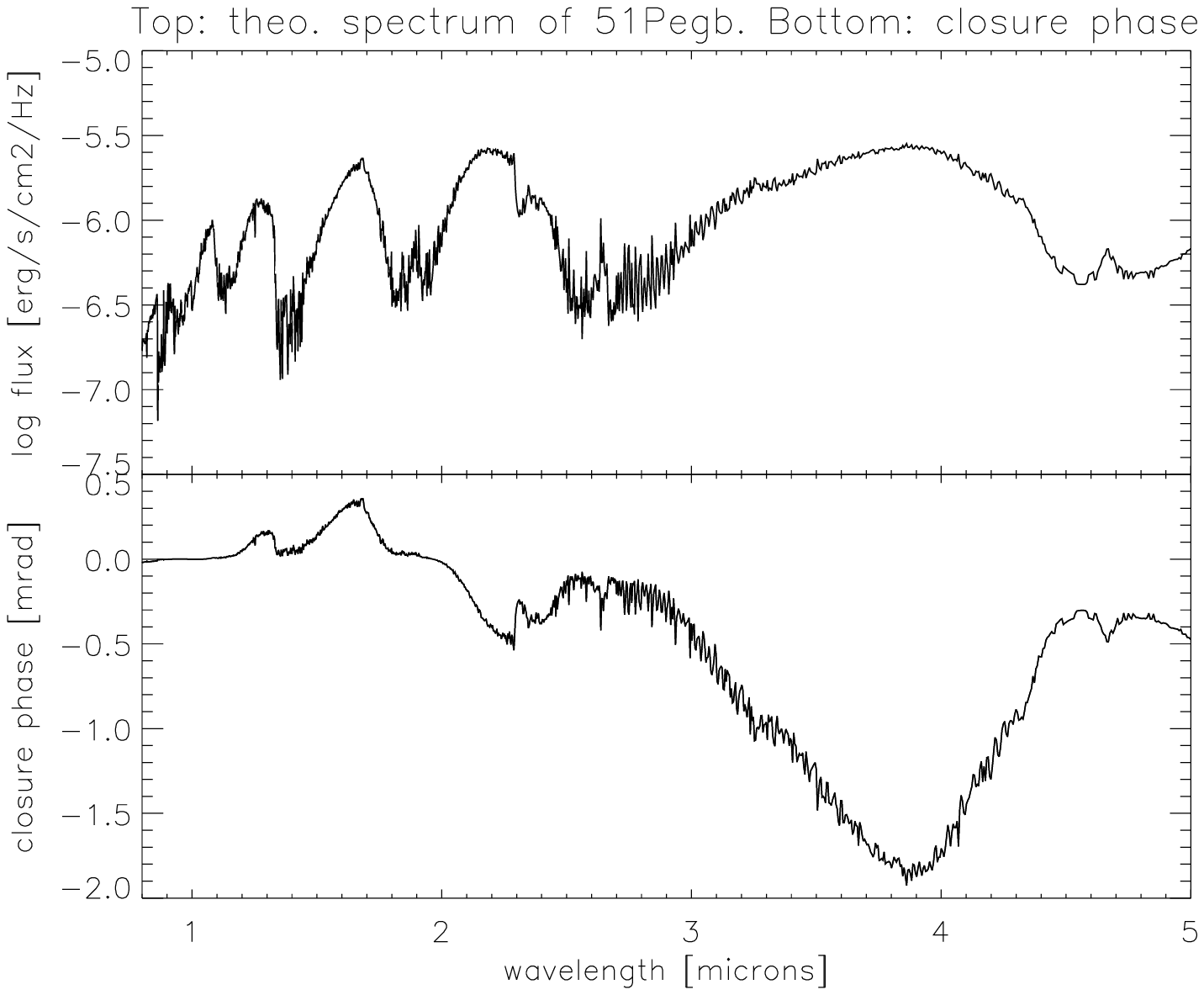}
   \end{tabular}
   \end{center}
   \caption[example]
   { \label{fig:clphase}
{\bf Spectral signal of planet in the closure phase.} \emph{Top panel:} Same as
for Fig.~\ref{fig:phase}. \emph{Bottom panel:} Closure phases in milliradian
calculated by us based on the theoretical spectrum of 51\,Peg\,B as well as of
the host star and a simulated observation with the near-IR instrument
AMBER/VLTI using the three telescopes UT1, UT3 and UT4 (baseline lengths102\,m,
62\,m, and 130\,m). It is evident that the spectrally resolved closure phases
contain a wealth of spectral information of the planet.}
   \end{figure}
%-------------

We have written a code to calculate simulated closure phases for known
extrasolar planetary systems, taking into account their theoretical spectra as
well as the geometry of the VLTI. We use theoretical spectra, which have been
calculated by Sudarsky, Burrows \& Hubeny (2003); see their paper for the
details on the underlying theory.

The complex visibility measured with a two-telescope interferometer for a
planetary system with the star on axis, and under the assumption that the
planet and the star are point sources, can be written as

\begin{equation}
\label{equ:vis}
V(u,v)=\frac{1+ r(\lambda) e^{-2 \pi i (\Delta \alpha u + \Delta \beta v)}}
            {1+r(\lambda)}~~,
\end{equation}

with $r(\lambda) = I_\mathrm{pl}/I_{\ast}$ being the contrast ratio between
planet and star, $\vec{s}=(\Delta \alpha, \Delta \beta)$ the separation vector
between star and planet in the tangent plane of the sky, and $u$ and $v$ the
spatial frequencies, i.e., the coordinates of the projected baseline vector in
units of the wavelength $\lambda$.

The phase measured between telescopes $i$ and $j$ is given by the argument of
(\ref{equ:vis}) and can be expressed as

\begin{eqnarray}
\label{equ:phase}
\Phi_{ij} & = & \arctan \left[ \frac{r(\lambda) \cdot \sin{(-2 \pi \Delta \alpha u + \Delta \beta v)}}
                      {1+r(\lambda) \cdot \cos{(-2 \pi \Delta \alpha u + \Delta \beta v)}} \right] \\
          & {~} & \nonumber\\
          & = & \arctan \left [\frac{r(\lambda) \cdot \sin{(\frac{-2 \pi}{\lambda} ~ \vec{s} \cdot \vec{B}_{ij})}}
                      {1+r(\lambda) \cdot \cos{(\frac{-2 \pi}{\lambda} ~ \vec{s} \cdot \vec{B}_{ij})}} \right]~~.
\end{eqnarray}
The closure phase for an array of three telescopes is calculated
by summing $\Phi_{ij}$ for the three baselines:
\begin{equation}
\label{equ:clphase}
\Phi_\mathrm{cl} = \Phi_{12} + \Phi_{23} + \Phi_{31}~~.
\end{equation}

We have retrieved the contrast ratio planet/star $r(\lambda)$ from the web site
of A.\ Burrows. For the separation $|s|$ between star and planet the published
semi-major axis is taken, but the orientation of $\vec{s}$ is unknown, and an
arbitrary value was taken for the purpose of the simulations. The projected
baselines $\vec{B}_{ij}$ are calculated for the VLTI 8\,m unit telescopes
(UTs).

\begin{figure}[t]
   \begin{center}
   \begin{tabular}{c}
   \includegraphics[height=11.8cm]{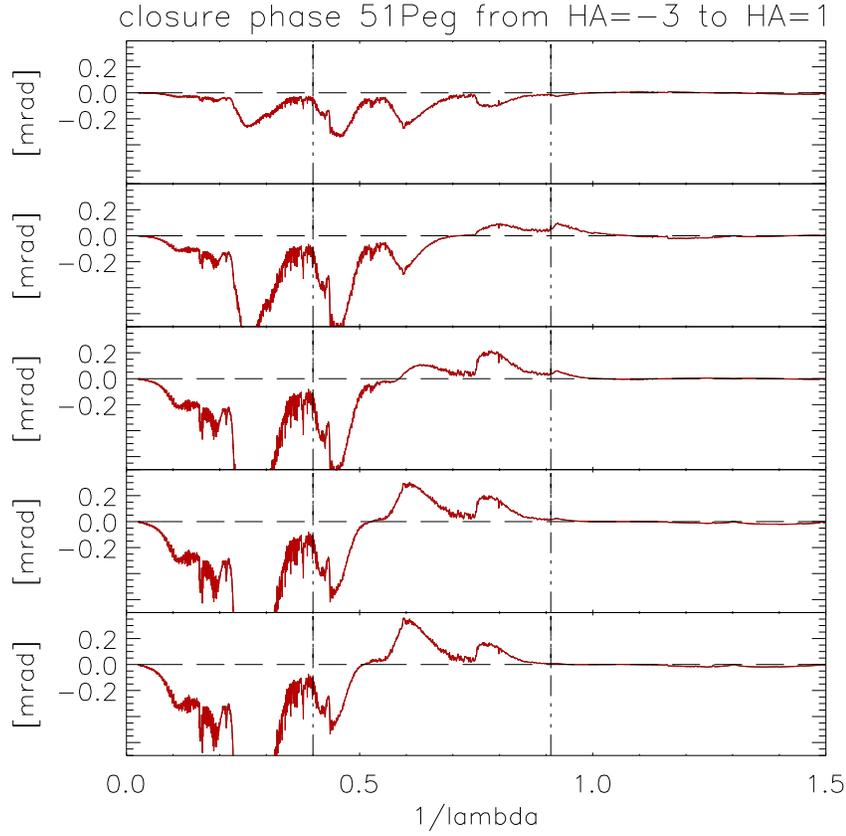}
   \end{tabular}
   \end{center}
   \caption[example]
   { \label{fig:rotsynt}
   {\bf Closure phase time series for the planet system 51\,Peg.}
   Simulations for observations with the VLTI array UT1, UT3, UT4
   for hour angles HA\,=\,$-3$, $-2$, $-1$, 0, +1\,hr (from top to bottom).
   Vertical lines indicate the wavelength range covered by AMBER
   (the near-IR bands between 1.1\,$\mu$m and 2.5\,$\mu$m).
   }
   \end{figure}

Figures~\ref{fig:phase} and \ref{fig:clphase} show in the top panels the
theoretical spectrum of the giant irradiated planet orbiting the solar-like
star 51\,Peg (Sudarsky, Burrows \& Hubeny 2003). The most prominent features
are CO and H$_2$O absorption bands in the near-IR. The bottom panel of
Fig.~\ref{fig:phase} displays the modeled phase for the single-baseline between
the two telescopes UT1 and UT3 (baseline length 102\,m) based on the shown
theoretical spectrum of 51\,Peg\,B as well as of the spectrum of the host star
(not shown). However, due to atmospheric turbulence, the absolute phase is not
an accessible quantity without a dual-beam facility allowing for phase
referencing. The bottom panel of Fig.~\ref{fig:clphase} shows the modeled
closure phases of the 51\,Peg system for the VLTI array comprised by the three
telescopes UT1, UT3 and UT4 (baseline lengths 102\,m, 62\,m, and 130\,m). It
can be clearly seen that the spectrally resolved phases and closure phases
contain a wealth of spectral information of the planet. Further results of the
simulations are shown in Fig.~\ref{fig:nulls} and Fig.~\ref{fig:rotsynt} and
are discussed later.

%%%%%%%%%%%%%%%%%%%%%%%%%%%%%%%%%%%%%%%%%%%%%%%%%%%%%%%%%%%%%
\section{NULLS IN THE CLOSURE PHASE}

%-------------
   \begin{figure}[t]
   \begin{center}
   \begin{tabular}{c}
   \includegraphics[height=12cm,angle=90]{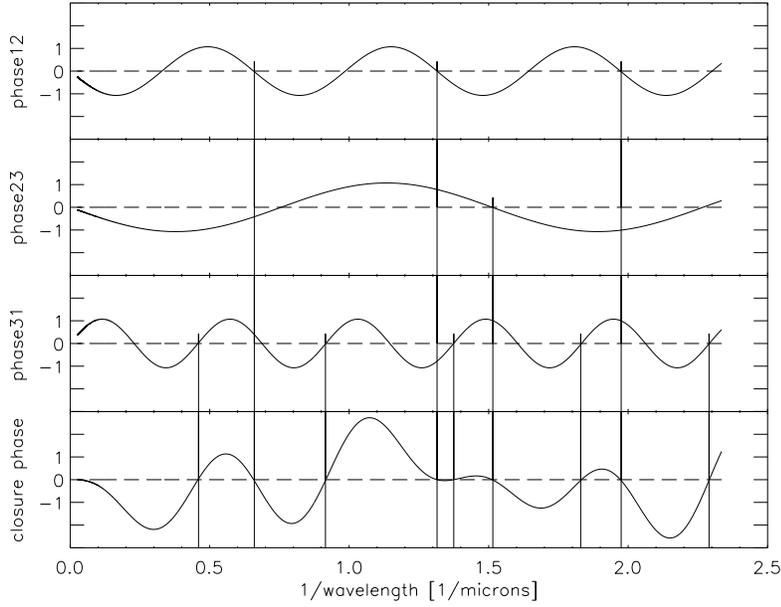}
   \end{tabular}
   \end{center}
   \caption[example]
   { \label{fig:nulls}
   {\bf Relation between the closure phase and the
   corresponding single-baseline phases.}
   The figure shows the simulated phase measurements
   for a system with a 3 milliarcsec separation at zenith
   observed with the VLTI telescopes UT1, UT3 and UT4
   and a constant flux ratio between the two components
   of 10$^{-3}$.
   The three top panels display the phases measured
   for the three single-baselines between telescope 1 and 2,
   2 and 3, and 3 and 1. (Note that they cannot be measured without
   a phase reference (PRIMA) and are plotted here only for
   theoretical considerations.)
   The bottom panel shows the simulated closure phase for
   these three telescopes. All phases are in millirad.
   It is obvious from this figure that the nulls in the
   closure phase are closely related to
   the nulls of the single-baselines phases: every second
   null of a single-baseline phase is also a null in the
   closure phase.
   }
   \end{figure}
%-------------

%-------------
%-------------

%-------------
   \begin{figure}[t]
   \begin{center}
   \begin{tabular}{c}
   \includegraphics[width=.48\linewidth,angle=0]{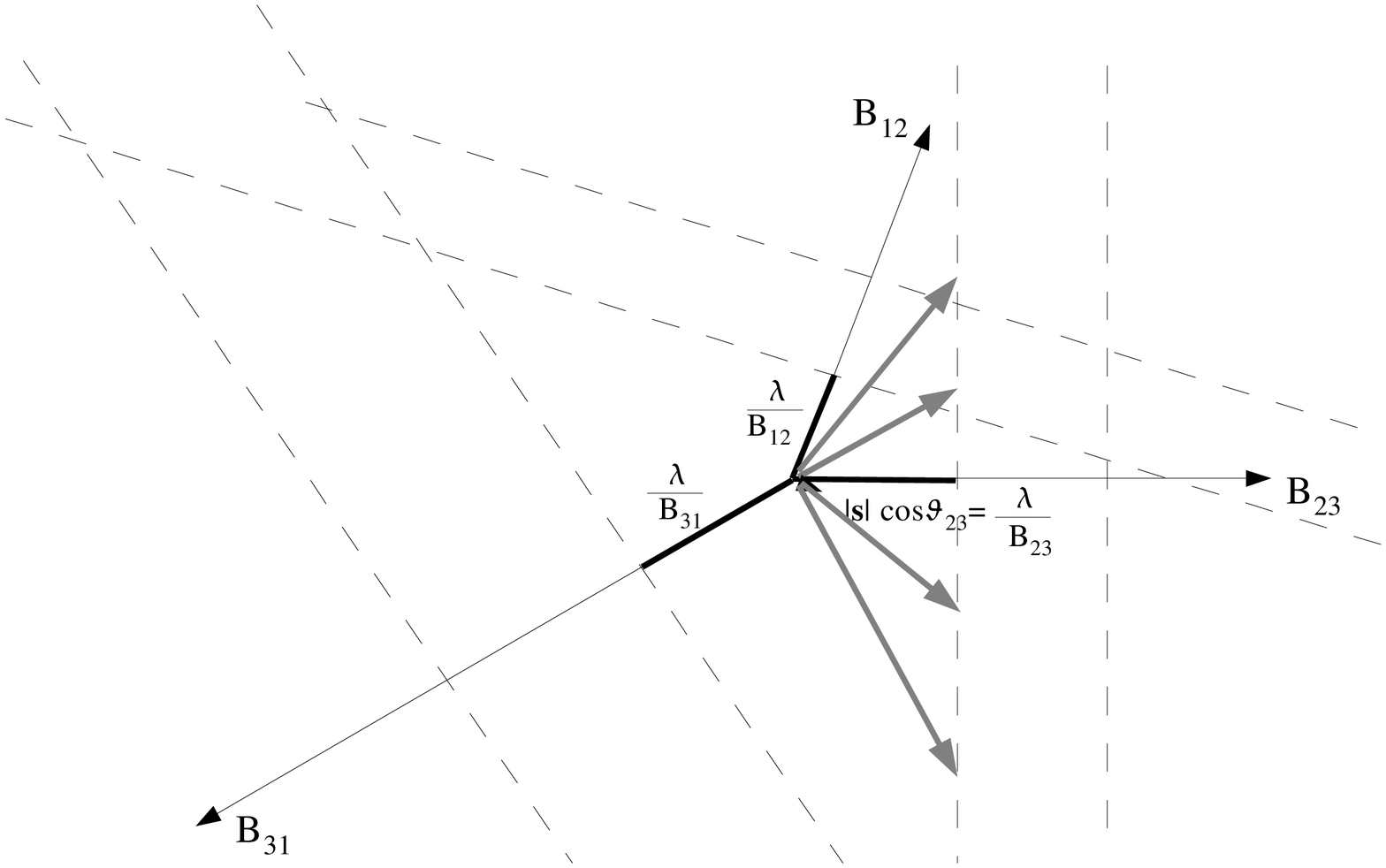}
   \includegraphics[width=.48\linewidth,angle=0]{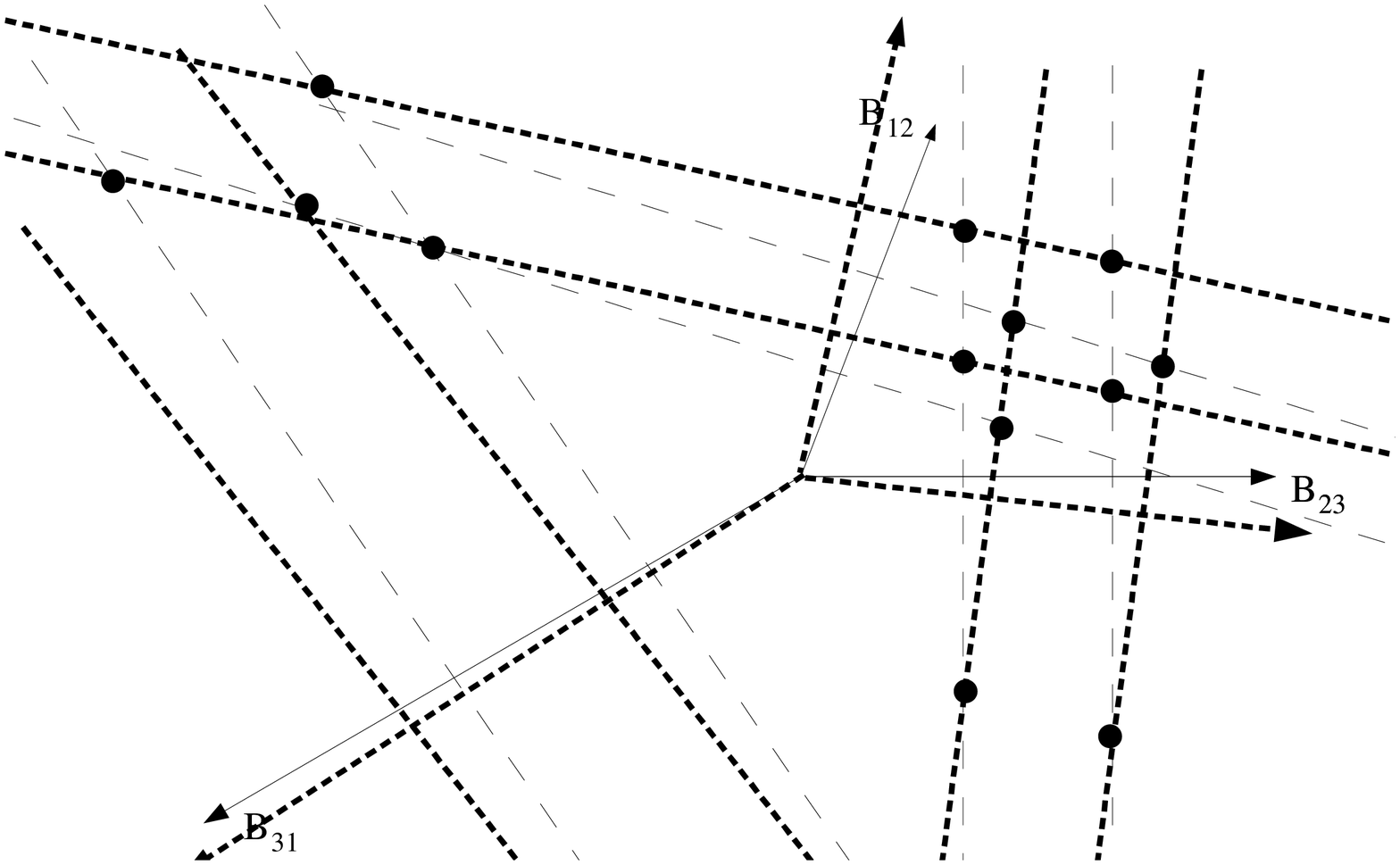}
   \end{tabular}
   \end{center}
   \caption[example]
   { \label{fig:liniennetz}
   {\bf Possible geometries of a star -- planet system as derived from nulls of
   the closure phase.}
   \emph{Left:} Determination of the nulls in the closure phase
   signal of a planetary system at a certain hour angle is a measure for
   the projection of the separation vector $\vec{s}$ between
   star and planet onto one of the three projected baselines.
   Due to the unknown orientation $\vartheta_{ij}$ we do not know onto
   which one. The dashed lines indicate the possible locations of $\vec{s}$
   for the nulls corresponding to $n=1$ and 2.
   \emph{Right:} Observations at a later hour angle gives another set of lines.
   The possible locations for $\vec{s}$ are now reduced to several grid points
   as marked by the intersections with the former lines. With a third
   observation, it will be possible to select the correct intersection point.
   }
   \end{figure}
%-------------

In Fig.~\ref{fig:nulls} we have plotted the simulated phases for two point
sources with a contrast ratio of 10$^{-3}$ and a separation of about 3
milliarcsec for the VLTI array comprised of the telescopes UT1, UT3 and UT4.
For the purpose of studying the relation between the closure phase and the
corresponding individual phases without the complication of the complex
spectral features of the source, the contrast ratio was chosen to be constant
for all wavelengths (``flat spectrum''). It can be seen from the plot that the
nulls in the closure phase are always also nulls in one of the corresponding
single-baseline phases. Furthermore, every second null of a single-baseline
phase is also a null of the closure phase.

The condition for single-baseline phases $\Phi_{ij}$ to be zero is that the dot
product between separation vector and projected baseline ($\vec{s} \cdot
\vec{B}_{ij}$) equals a multiple of $\lambda/2$. The nulls in the
single-baseline phase depend only on the interferometer geometry
($\vec{B}_{ij}$) and the planetary system geometry ($\vec{s}$), i.e.,
separation and orientation of the system, but are independent of the planet and
star spectra ($r(\lambda$)).

In order to derive the condition for nulls in the closure phase, we insert the
expression for $\Phi_{ij}$ (Eqn.~\ref{equ:phase}) into the closure phase
relation (Eqn.~\ref{equ:clphase}) and set the denominators to 1. The latter
approximation is legitimate since the contrast ratio between planet and star
can be assumed to be much smaller than 1. (We note that this assumption is not
necessary for the derivation of the conditions for the nulls in the closure
phase. For simplicity we make it nevertheless.) The closure phase then becomes
\begin{equation}
\label{equ1}
\Phi_\mathrm{cl} = \arctan \left[ r \sin{(-\frac{2 \pi}{\lambda} ~ \vec{s} \cdot \vec{B}_{12})} \right]
                 + \arctan \left[ r \sin{(-\frac{2 \pi}{\lambda} ~ \vec{s} \cdot \vec{B}_{23})} \right]
                 + \arctan \left[ r \sin{(-\frac{2 \pi}{\lambda} ~ \vec{s} \cdot \vec{B}_{31})} \right]~~.
\end{equation}
Taking the closure condition that the baseline vectors add up to zero into
account ($\vec{B}_{12} + \vec{B}_{23} + \vec{B}_{31} = 0$), and making the
substitutions
\begin{eqnarray}
\gamma & := & -\frac{2 \pi}{\lambda} ~ \vec{s} \cdot \vec{B}_{12} \\
p & := & \frac{\vec{s} \cdot \vec{B}_{31}}{\vec{s} \cdot \vec{B}_{12}}
\end{eqnarray}
we can rewrite Eqn.~\ref{equ1} as
\begin{equation}
\label{equ2}
\Phi_\mathrm{cl} = \arctan [~r \sin{(\gamma})~]
                 - \arctan [~r \sin{(\gamma)} \cos{(\gamma p)}+r \cos{(\gamma)} \sin{(\gamma p)}~]
                 + \arctan [~r \sin{(\gamma p)}~]~~.
\end{equation}

From Eqn.~\ref{equ2} we can derive the conditions for nulls of the closure phase.
The closure phase $\Phi_\mathrm{cl}$ becomes zero
if $\gamma$ equals a multiple of $2 \pi$:
\begin{equation}
\Phi_\mathrm{cl} = - \arctan [~r \sin{(\gamma p)}~]
                   + \arctan [~r \sin{(\gamma p)}~] = 0~~.
\end{equation}

The condition $\gamma = n \cdot 2 \pi$ is equivalent to $\vec{s} \cdot
\vec{B}_{12} = n \cdot \lambda$. Comparing that with the condition for a null
in the single-baseline phase $\Phi_{12}$, namely $\vec{s} \cdot \vec{B}_{12} =
n \cdot \lambda / 2$, shows that for every second null in $\Phi_{12}$ the
closure phase also has a null.

Furthermore, it can be seen from Eqn.~\ref{equ2} that for $\gamma \cdot p = n
\cdot 2 \pi$, or equivalently $\vec{s} \cdot \vec{B}_{31} = n \cdot \lambda $,
the closure phase has also a null corresponding to the baseline $B_{31}$. A
third condition for a null in $\Phi_\mathrm{cl}$ exists for baseline B$_{23}$
and it can be derived from Eqn.~\ref{equ2} by choosing another appropriate
substitution.

To summarize, the closure phase has a null when the dot product between
separation vector $\vec{s}$ and baseline vector $\vec{B}_{ij}$ equals multiples
of $\lambda$:

\begin{equation}
\label{equ:null} \vec{s} \cdot \vec{B}_{ij} = n \cdot \lambda \quad
\Longleftrightarrow \quad |s| \cdot |B_{ij}| ~ \vartheta_{ij} = n \cdot
\lambda~~,
\end{equation}

with $\vartheta_{ij}$ being the angle between $\vec{s}$ and $\vec{B}_{ij}$.

This completes our proof of the relation between nulls in the closure phase and
in the phases on the individual baselines illustrated in Fig.~\ref{fig:nulls}.
It also shows that the nulls in the closure phase are independent of the
planet/star contrast ratio and do only depend on the system geometry and the
interferometer geometry.

%%%%%%%%%%%%%%%%%%%%%%%%%%%%%%%%%%%%%%%%%%%%%%%%%%%%%%%%%%%%%
\section{EARTH ROTATION SYNTHESIS}

We now proceed to describe an algorithm to determine the spectrum of a planet
and the geometry of the star -- planet system (i.e., the angular separation and
position angle of the planet with respect to the star at the time of the
observation) from closure phase observations, without any a priori knowledge of
these quantities. The stellar spectrum (or more precisely, the composite
spectrum of the star -- planet system) can be presumed to be known, however,
since this is easily measurable with a simple spectrograph. Determining the
planet spectrum is therefore equivalent to measuring the contrast ratio
$r(\lambda)$.

As we have seen in the previous section, the nulls of the closure phase contain
information about the system geometry, because they are related to the nulls of
single-baseline phases. It is not known a priori, however, to which single
baselines the individual nulls in the closure phase correspond. We can thus
interpret Eqn.~\ref{equ:null} as follows:

By determining a wavelength for which the closure phase is zero, we have a
measure for the projection $|s| \cdot \vartheta_{ij}$ of the separation vector
$\vec{s}$ onto one of the individual baselines $\vec{B}_{ij}$, but we do not
know onto which one. Furthermore, we do not necessarily know the order $n$ of
the null. The geometrical locus of all possibilities for $\vec{s}$ is therefore
a set of straight lines, as shown in the left panel of
Fig.~\ref{fig:liniennetz}; each line in this figure is perpendicular to one of
the baselines and corresponds to an assumed set of $\{i,j,n\}$. (For clarity,
only the lines corresponding to $n = 1 $ and $n = 2$ are shown.)

The set of baselines vectors formed by the three telescopes changes with time
due to the Earth's rotation; a second observation therefore gives an
independent constraint on the geometry as shown in the right panel of
Fig.~\ref{fig:liniennetz}. The possible system geometries are now the
intersections of the set of lines corresponding to the first observation with
the set corresponding to the second. This discrete set of points has been
marked with filled circles in the figure. A third observation will produce yet
another independent set of lines. In general, this set will pass through only
one of the previously marked points: this is the true separation
vector.\footnote{One may be concerned that the infinite number of lines
corresponding to $n = 1, 2, 3, \ldots$ for each observation and each baseline
may invalidate this argument. It should be noted, however, that arbitrarily
high values of $n$ can be excluded because then the nulls would be densely
spaced in $\lambda$. If $n$ is large, one would thus also observe the nulls of
order $n-1$ or $n+1$ within the bandpass.} It is thus possible to derive the
system geometry unambiguously with three observations.

Once the separation vector $\vec{s}$ is known, it is straightforward to
determine $r(\lambda)$ through a numerical inversion of Eqn.~\ref{equ2},
separately for each $\lambda$. The only difficulty occurs at the nulls, where
$r$ cannot be constrained. But since three observations are needed in any case
for the determination of $\vec{s}$, one can perform the three inversions and
combine them with (wavelength-dependent) weights appropriate for their
respective signal-to-noise ratios. Consider, for example, the wavelength range
around $\lambda = 1.67\,\mu$m ($1 / \lambda = 0.6\,\mu$m$^{-1}$) in
Fig.~\ref{fig:rotsynt}. The observation at $-1$\,hr (third panel) gives very
low signal-to-noise close to the null near $1 / \lambda = 0.6\,\mu$m$^{-1}$,
but observations at other times can be used to infer the planet spectrum in
this wavelength region.

%%%%%%%%%%%%%%%%%%%%%%%%%%%%%%%%%%%%%%%%%%%%%%%%%%%%
\section{OUTLOOK}

We have shown that the separation vector and spectrum of extrasolar planets can
be determined from closure phase measurements in a deterministic way, with a
non-iterative algorithm, and without any a priori assumptions. For a practical
application of this technique, a few additional complications will have to be
considered:
\begin{itemize}
\item{The assumption that the star and the planet are point sources will have
to be relaxed. Taking into account that the star is slightly resolved by the
interferometer in Eqn.~\ref{equ:vis} complicates the analysis and implies that
the relation between nulls in the closure phase and the single-baseline phases
is only approximately fulfilled.}
\item{For observations from the ground, the useful wavelength range in the
near-infrared is limited to the atmospheric windows and thus non-contiguous.}
\item{The finite signal-to-noise ratio of realistic observations will lead to
an uncertainty in determining the exact wavelengths of closure phase nulls. The
lines and intersection points of Fig.~\ref{fig:liniennetz} will thus be
broadened.}
\item{The non-zero time needed to accumulate sufficient signal-to-noise on the
closure phase means that the interferometer geometry will change slightly
during the observations. This is again equivalent to a slight broadening of the
allowed regions in Fig.~\ref{fig:liniennetz}.}
\item{If the time required to accumulate all observations is not short compared
to the orbital period of the planet, the motion of the planet will also
contribute an uncertainty in the derived geometry.}
\end{itemize}
These practical complications will certainly be at least partially compensated
by a larger number of observations; one would probably take ten or more rather
than the minimum three. One could also combine the algorithm presented here
with a global $\chi^2$ minimization, in which the orbital parameters and
spectrum of the planet are fitted to the observations. The purpose of our
algorithm would then consist of providing a robust starting point for the
$\chi^2$ minimization algorithm, which is usually crucial to ensure convergence
to the correct minimum.

The AMBER instrument has arrived on Cerro Paranal this year and the first
commissioning runs have taken place. It is now necessary to determine if the
necessary closure phase precision (better than 0.1\,mrad, see
Fig.~\ref{fig:clphase}) can be reached. If so, the VLTI will provide
unprecedented opportunities for observations of extrasolar planets and their
spectra.

%%%%%%%%%%%%%%%%%%%%%%%%%%%%%%%%%%%%%%%%%%%%%%%%%%%%%%%%%%%%%
\acknowledgments     %>>>> equivalent to \section*{ACKNOWLEDGMENTS}

We are grateful to our colleagues at the Sterrewacht Jeff Meisner, Bob Tubbs
and Walter Jaffe for helpful discussions on the topic of this article. VJ
acknowledges support by a Marie Curie Fellowship of the European Community
program ``Structuring the European Research Area'' under contract number
FP6-501875.

%%%%%%%%%%%%%%%%%%%%%%%%%%%%%%%%%%%%%%%%%%%%%%%%%%%%%%%%%%%%%
%%%%% References %%%%%

%\bibliography{spie_viki_bib}   %>>>> bibliography data in report.bib

\begin{thebibliography}{14}

\bibitem{a1x}{Akeson, R.L., \& Swain, M.R.\ 1999.\ {\it Differential phase mode
with the Keck Interferometer.} In {\it Working on the fringe: optical and IR
interferometry from ground and space.} Ed.\ S.\ Unwin \& R.\ Stachnik, Proc.\
ASP Conf.\ Vol.\ 194, p.\ 89}

\bibitem{c1x}{Charbonneau, D., Brown, T.M., Noyes, R.W., \& Gilliland, R.L.\
2002.\ {\it Detection of an Extrasolar Planet Atmosphere}, ApJ 568:377}

\bibitem{j1x}{Jennison, R.C.\ 1958.\ {\it A phase sensitive interferometer
technique for the measurement of the fourier transforms of spatial brightness
distributions of small angular size}, MNRAS 118:276}

\bibitem{l1x}{Lopez, B., \& Petrov, R.G.\ 2000.\ {\it Direct Detection of Hot
Extrasolar Planets Using Differential Interferometry.} In {\it From Extrasolar
Planets to Cosmology: The VLT Opening Symposium.} Ed.\ J.\ Bergeron \& A.\
Renzini, p.\ 565}

\bibitem{m3x}{Malbet, F., Driebe, T.M., Foy, R., Fraix-Burnet, D., Mathias, P.,
Marconi, A., Monin, J., Petrov, R.G., Stee, P., Testi, L., \& Weigelt, G.\ 2004
{\it Science program of the AMBER consortium.} In {\it New Frontiers in Stellar
Interferometry.} Ed.\ W.A.\ Traub, Proc.\ SPIE 5491, in press}

\bibitem{m1x}{Mayor, M., \& Queloz, D.\ 1995.\ {\it A Jupiter-Mass Companion to
a Solar-Type Star}, Nature 378:355}

\bibitem{m2x}{Mayor, M., Udry, S., Naef, D., Pepe, F., Queloz, D., Santos,
N.C., \& Burnet, M.\ 2003.\ {\it The CORALIE survey for southern extra-solar
planets, XII.\ Orbital solutions for 16 extra-solar planets discovered with
CORALIE}, A\&A 415:391}

\bibitem{m4x}{Meisner, J.\ 2004.\ {\it Direct detection of exoplanets using
long-baseline interferometry and visibility phase.} In {\it Extrasolar planets:
today and tomorrow.} Ed.\ J.P.\ Beaulieu, A.\ Lecavelier des Etangs, \& C.\
Terquem, ASP Conf.\ Ser., in press}

\bibitem{p1x}{Petrov, R.G., Malbet, F., Weigelt, G., Lisi, F., Puget, P.,
Antonelli, P., Beckmann, U., Lagarde, S., Lecoarer, E., Robbe-Dubois, S., et
al.\ 2003.\ {\it Using the near infrared VLTI instrument AMBER.} In {\it
Interferometry for Optical Astronomy II.} Ed.\ W.A.\ Traub, Proc.\ SPIE 4838,
p.\ 924}

\bibitem{q1x}{Quirrenbach, A.\ 2000.\ {\it Astrometry with the VLT
Interferometer.} In {\it From extrasolar planets to cosmology: the VLT opening
symposium.} Ed.\ J.\ Bergeron, \& A.\ Renzini, A., p.\ 462}

\bibitem{q2x}{Quirrenbach, A., \& Mariotti, J.M.\ 1997.\ {\it The VLTI and the
universe: conference summary.} In {\it Science with the VLT Interferometer.}
Ed.\ F.\ Paresce, p.\ 339}

\bibitem{s1x}{Segransan, D.\ 2001.\ {\it The very low mass stars of the solar
neighborhood: multiplicity and mass-luminosity relations.} PhD thesis,
Grenoble}

\bibitem{s2x}{Sudarsky, D., Burrows, A., \& Hubeny, I.\ 2003.\ {\it Theoretical
spectra and atmospheres of extrasolar giant planets}, ApJ 588:1121}

\bibitem{v1x}{Vannier, M., Petrov, R.G., Schoeller, M., Lopez, B., Antonelli,
P., Lagarde, S., Robbe-Dubois, S., Morel, S., \& Rantakyro, F.\ 2004.\ {\it
Design and tests for the correction of atmospheric and instrumental effects on
color-differential phase with AMBER/VLTI.} In {\it New Frontiers in Stellar
Interferometry.} Ed.\ W.A.\ Traub, Proc.\ SPIE 5491, in press}

\bibitem{v2x}{Vidal-Madjar, A., D\'esert, J.M., Lecavelier des Etangs, A.,
H\'ebrard, G., Ballester, G.E., Ehrenreich, D., Ferlet, R., McConnell, J.C.,
Mayor, M., \& Parkinson, C.D.\ 2004.\ {\it Detection of oxygen and carbon in
the hydrodynamically escaping atmosphere of the extrasolar planet
HD\,209458\,b}, AJ 604:L69}


\end{thebibliography}
%\bibliographystyle{spiebib}   %>>>> makes bibtex use spiebib.bst

\end{document}